\documentclass[pre,aps,twocolumn,showpacs,floatfix,superscriptaddress]{revtex4-1}
\usepackage{graphicx}
\usepackage{amsmath,amssymb}
\usepackage{mathtools}
\usepackage{dcolumn}
\usepackage{bm}
\usepackage{subfigure}
\usepackage{epsfig}
\usepackage{afterpage}
\usepackage[colorinlistoftodos]{todonotes}
\usepackage[colorlinks=true, allcolors=blue]{hyperref}

\newcommand{\bea}{\begin{eqnarray}}
\newcommand{\eea}{\end{eqnarray}}
\newcommand{\beq}{\begin{equation}}
\newcommand{\eeq}{\end{equation}}
\newcommand{\be}{\begin{equation}}
\newcommand{\ee}{\end{equation}}
\newcommand{\beqa}{\begin{eqnarray}}
\newcommand{\eeqa}{\end{eqnarray}}

\begin{document}

\title{Exact Hyperuniformity Exponents and Entropy Cusps in \\Models of Active-Absorbing Transition}
\author{Rahul Dandekar} 
\affiliation{The Institute of Mathematical Sciences,
C.I.T. Campus, Taramani, Chennai-600113, India} 
\affiliation{Homi Bhabha National Institute, Training School Complex, Anushakti Nagar, Mumbai-400094, India}

\begin{abstract}
Recent studies of nonequilibrium phase transitions have shown that in many systems which have transitions involving an arrested phase, the arrested states show suppressed density fluctuations and a cusp in the configurational entropy at the transition point. We study quasistatic driving in the 1D fixed-energy abelian sandpile model, and in conserved lattice gases with range $n$ in 1D, and exactly determine the measure over the absorbing states for both cases, along with the behaviour of density fluctuations and the configurational entropy of absorbing states. We show that both models exhibit hyperuniformity near the transition, and also that the configurational entropy shows a cusp at the transition point in the conserved lattice gases.
\end{abstract}

\maketitle

\emph{Introduction} Models of active-absorbing phase transitions \cite{lubeck2004universal,henkel2008non} exhibit a change in the dynamical behaviour of the system from a phase where after a finite time the dynamics is arrested or dies, falling into what are called absorbing states, to a phase where a steady state with finite activity is reached in the thermodynamic limit, as a parameter $p$ is varied across the transition point $p_c$. For $p< p_c$, if there are infinitely many absorbing states, the measure over them can depend on the initial conditions. The measurement of near-critical properties in the absorbing phase is thus complicated by the necessity of defining a measure over the absorbing states. The construction of a `natural' measure over absorbing states is an important issue in current debates over the universality class of models with a conserved field \cite{basu2012fixed,le2015exact,dickman2015particle}. It should be noted that in models of self-organized criticality the presence of an infinitesimal drive and boundary dissipation leads to a unique measure over absorbing states that can, in many cases, be exactly determined\cite{dhar1999abelian}, and such a measure can exhibit hyperuniformity that is analytically tractable \cite{grassberger2016oslo}.

In this letter, we study two models with a conserved density that show an active-absorbing phase transition in one dimension. The first model is the abelian sandpile model (ASM), while the second is the class of conserved lattice gases with extended range (CLGs) whose the active steady state behaviour was exactly solved in \cite{dandekar2013class}. We show that the introduction of a quasistatic drive leads to an exactly solvable measure on the absorbing states in these models. In glassy and jammed systems, such a drive corresponds either to very slow external driving, or to the limit $T \rightarrow 0$, and is a natural way to study arrested states in these systems \cite{radjai2002turbulentlike,howell1999stress,maloney2006amorphous}. The external drive we introduce does not break any symmetries of the original dynamics, and allows us to characterize the near critical properties of the absorbing states exactly.

Two interesting characterizations of absorbing or arrested states that have attracted attention in recent years are the hyperuniformity of particle number fluctuations and the configurational entropy.

Translationally invariant arrangements of points in space, or occupied sites on a lattice, can be classified on a scale from regular to random based on the scaling of the number fluctuations in a large probe length scale $l$. For a periodic pattern, the number fluctuations scale as the surface area of the probe volume, $l^{d-1}$. In contrast, when particles are arranged randomly in space, the number fluctuations scale as $l^{d/2}$. In general, if number fluctuations scale as $l^{\alpha}$ as $l\rightarrow \infty$ with $\alpha$ between $\frac{d}{2}$ and $d-1$, the system is said to exhibit hyperuniformity\cite{hexner2015hyperuniformity,torquato2016hyperuniformity,torquato2003local,ghosh2017fluctuations}. Several studies have found hyperuniformity in jammed packing and athermal structures \cite{atkinson2016critical,weijs2015emergent,tjhung2015hyperuniform,gradenigo2015edwards,hexner2018two}. Numerical simulations of models of active-absorbing phase transitions with a conserved density have been shown to have hyperuniform fluctuations at the critical point \cite{hexner2015hyperuniformity, hexner2017noise}. Away from the critical point, hyperuniformity is seen only over a finite length scale, the divergence of which as the critical point is approached defines the hyperuniformity correlation exponent $\nu_h$. The exponents $\nu_h$ and $\alpha$ are believed to depend only on the universality class, but so far there are no models for which they can be exactly determined.

\begin{figure}
	\centering
	\includegraphics[width=0.8\columnwidth]{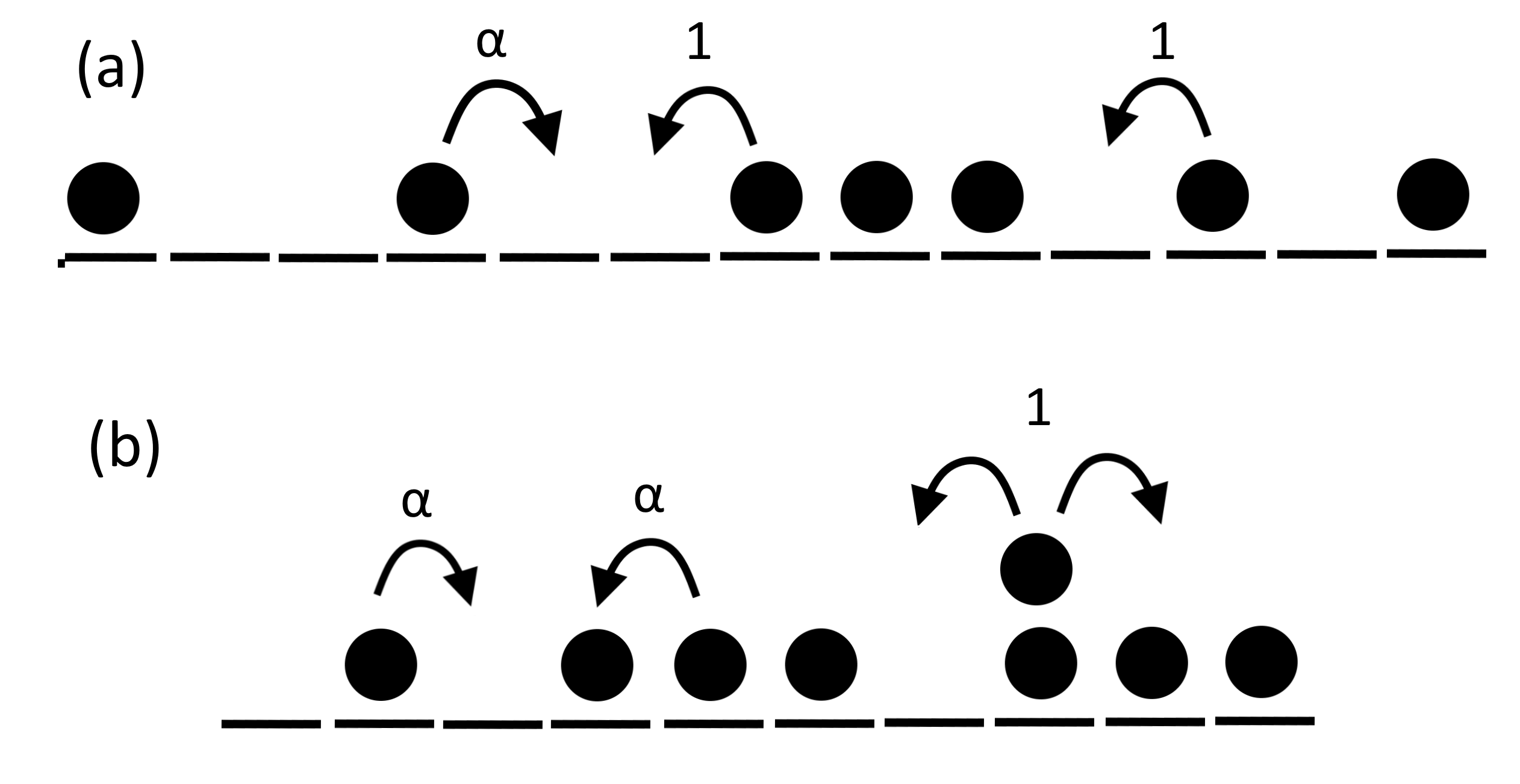}
	\caption{The transition rates for the models studied in this Letter. (a) The CLG with range $n=2$ in a transience field of strength $\alpha$. Setting $\alpha>0$ allows isolated particles, such as the second particle from the left, to move, reactivating the absorbing configurations. (b) The ASM driven with an field of strength $\alpha$. When $\alpha=0$ only the toppling moves for unstable sites, in which one particle is transferred to either side, are allowed.}
	\label{clg2}
\end{figure}

Martiniani et al. \cite{martiniani2019quantifying} measured the behaviour of a measure of configurational entropy, the computable information density (CID), for various active-absorbing phase transitions and found that the behaviour of the entropy of a typical configuration generically shows a cusp at the transition. Other studies since have extended this result to other measures of configurational entropy and more systems \cite{avinery2019universal,zu2019information}. However, the existence of a cusp has not previously been shown exactly for any model. 

We show that for the Abelian Sandpile Model in 1D and Conserved Lattice Gases with extended range, the measure over absorbing states obtained under quasistatic drive exhibits hyperuniformity. We analytically determine the hyperuniformity exponents, and demonstrate exactly the existence of an entropy cusp at the transition point.


\emph{1D Conserved Lattice Gases with extended range} Conserved Lattice gases are models of particles on the lattice with exclusion interaction, and the property that particles can only move if another particle is present within a specified range. The class of conserved lattice gases we study in this section was defined in \cite{dandekar2013class}, where the properties of the active steady state were characterised exactly. Consider $N$ particles on $L$ sites on a 1D ring, such that each site can have no more than 1 particle. The notation $1^{n_1} 0 ^{n_2} 1^{n_3} 0^{n_4}\dots$ denotes a configuration with a contiguous cluster of $n_1$ particles, followed by a contiguous cluster of $n_2$ empty sites, and so on. The allowed transitions for the model with range $n$ are (all transitions happen at rate $1$)

\beqa
110^{k} &\rightarrow& 1010^{k-1} ~~ \mbox{for all} ~~ k>0 \nonumber\\
0^k11 &\rightarrow& 0^{k-1}101 ~~ \mbox{for all} ~~ k>0 \nonumber\\
1 0^{i} 1 0^{j} 1 &\rightarrow& 1 0^{i-1} 1 0^{j+1} 1 ~~ \mbox{if} ~~ i+j\le n \label{eq:1dclgrates}\\
1 0^{i} 1 0^{j} 1 &\rightarrow& 1 0^{i+1} 1 0^{j-1} 1 ~~ \mbox{if} ~~ i+j\le n\nonumber
\eeqa

For $n=1$, this is the model known as the Conserved Lattice Gas. It can be seen that for this model, clusters of $0$s of length $>1$ are transient, since they are not created by the dynamics, and their number reduces monotonically with time. Only for $\rho>\frac{1}{2}$, all transient clusters of $0$s vanish until there are only isolated $0$s in the system, while for $\rho<\frac{1}{2}$, the dynamics eventually enters an absorbing configuration with only isolated $1$s. 

In the model with range $n$,  if a particle is either right next to another particle, or the total number of $0$s immediately to the left and right is not greater than $n$, the particle is 'active', and hops to a neighbouring empty site at rate $1$. For the model with range $n$, clusters of $0$s of length $>n$ are transient. A active steady state exists for $\rho > \frac{1}{n+1}$, and consists of all configurations with no $0$ clusters of length $>2$, with equal weight. Using a generating function formalism, the properties of the active steady state were determined in \cite{dandekar2013class}. Denoting the grand canonical partition function by $C^{act}(x,y)$, where $x$ is the fugacity of particles and $y$ the fugacity of sites, it was shown that
\beqa
C^{act}_n(x,y) &=& \left(1 - x y \frac{1- y^n}{1-y} \right)^{-1}, \label{eq:1dclgactss}\\
\mbox{giving } \rho_a \sim (\rho-\rho_c)^{n} & &\mbox{and } \xi \sim (\rho-\rho_c)^{-1}, \nonumber
\eeqa
where $\rho_a$ is the density of active (moveable) particles at density $\rho$, and $\xi$ is the correlation length. Hence the critical exponents for the model with range $n$ are $\beta=n$ and $\nu_{\perp} = 1$.

\emph{Transience field} We now introduce a field $\alpha$ which simulates an external drive, with the limit $\alpha \rightarrow 0$ being equivalent to quasistatic driving. The field is defined by two additional rates
\beqa
0^{i}10^{j} &\rightarrow^{\alpha} 0^{i+1}10^{j-1} ~~ \mbox{for} ~~ k\ge n, \nonumber\\
0^{i}10^{j} &\rightarrow^{\alpha} 0^{i-1}10^{j+1} ~~ \mbox{for} ~~ k\ge n.
\eeqa
These reactions only apply to isolated particles which cannot move according to the rates in eqns. (\ref{eq:1dclgrates}). 

It is convenient to think of the transient dynamics of the CLG models as the system going down a 'transience ladder', with the height of the ladder being the total length of $0$ clusters of length $>n$. The usual CLG dynamics takes the system down the ladder one step when a hop occurs into one of these clusters, reducing its length by 1. The external field dynamics allows the system to take reverse steps up the 'transience ladder', at a rate $\alpha$. As $\alpha \rightarrow 0$, the system settles down on the lowest rung on the transience ladder available at that density. Earlier studies have introduced a field in active-absorbing models which couples to the activity \cite{lubeck2002mean,lubeck2002scaling}. However, a transience field couples to the position of the system on the transience ladder, and hence is relevant to understanding the nature of the absorbing and transient states.

For any $\alpha$, the dynamics obeys detailed balance, and the equilibrium measure is given by (the $n_i$s below can be zero)
\beqa
W(1 0^{n_1} 1 0^{n_2} 1 0^{n_3} \dots) = \alpha^{\sum_i (n_i-n) \Theta(n_i-n)} \label{eq:clgwts}
\eeqa
Defining $Z(N,L,\alpha)$ is the partition function of the system of $N$ particles on $L$ sites with field $\alpha$, we have the grand canonical partition function,
\beqa
C(x,y,\alpha) = \left(1 - x y \frac{1- y^n}{1-y} - \frac{\alpha x y^{n+1}}{1-\alpha y} \right)^{-1}. \label{eq:1dclggenfunc}
\eeqa

The grand canonical partition function allows us to analyse the system using methods from generating function theory \cite{wilf2005generatingfunctionology}, see Appendix A for details. Using the notation $\Lambda = y^{-1}$, let us denote the largest pole of $C(x,\Lambda^{-1},\alpha)$ as $\Lambda_*(x,\alpha)$. Then, in the thermodynamic limit, a fugacity $x$ is equivalent to a density
\beq
\rho = x \frac{\partial}{\partial x} \log{(\Lambda_*(x,\alpha))}.
\eeq

\begin{figure*}
\begin{subfigure}
	\centering
	\includegraphics[width=\columnwidth]{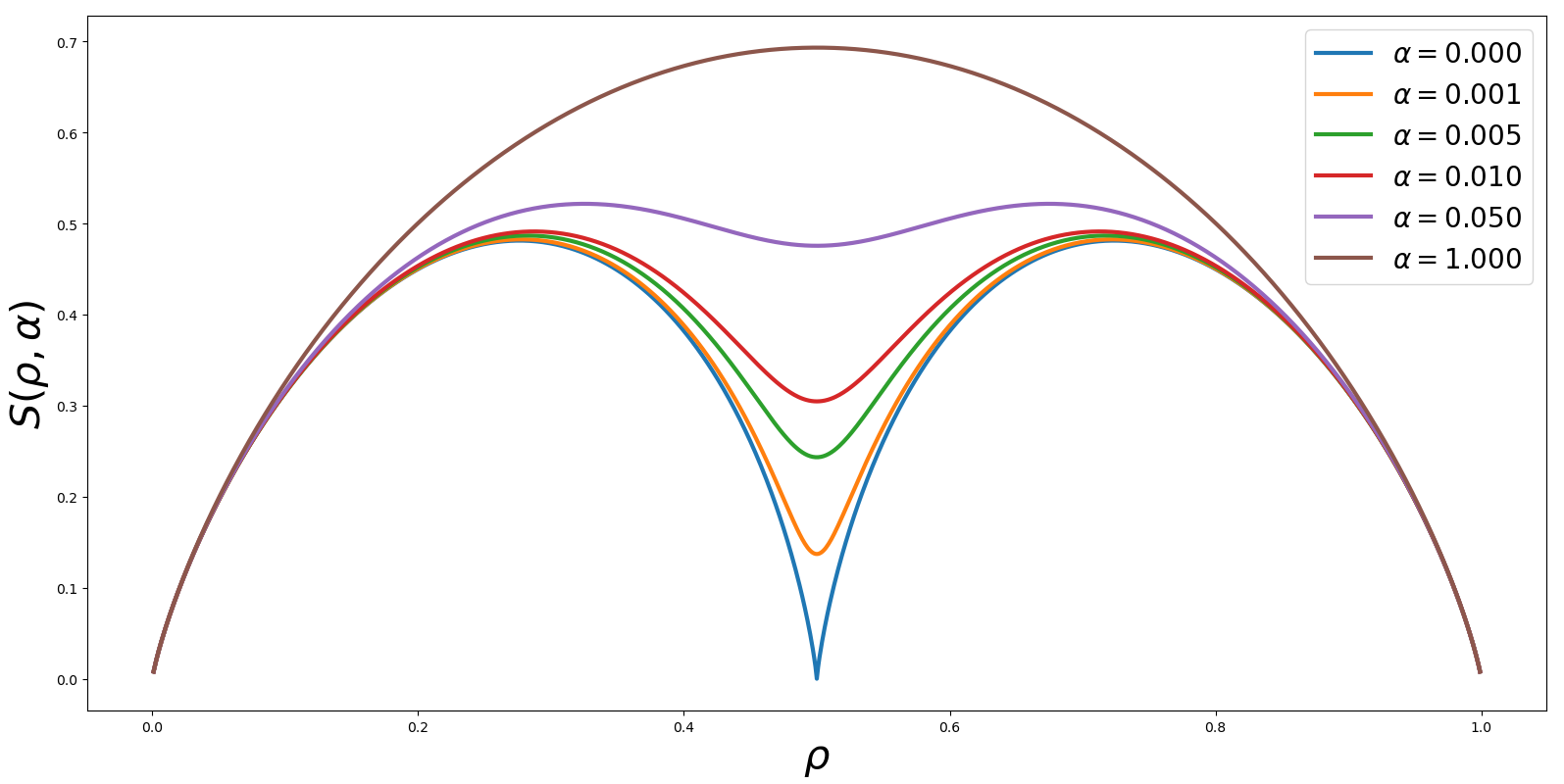}
\end{subfigure}
\begin{subfigure}
	\centering
	\includegraphics[width=\columnwidth]{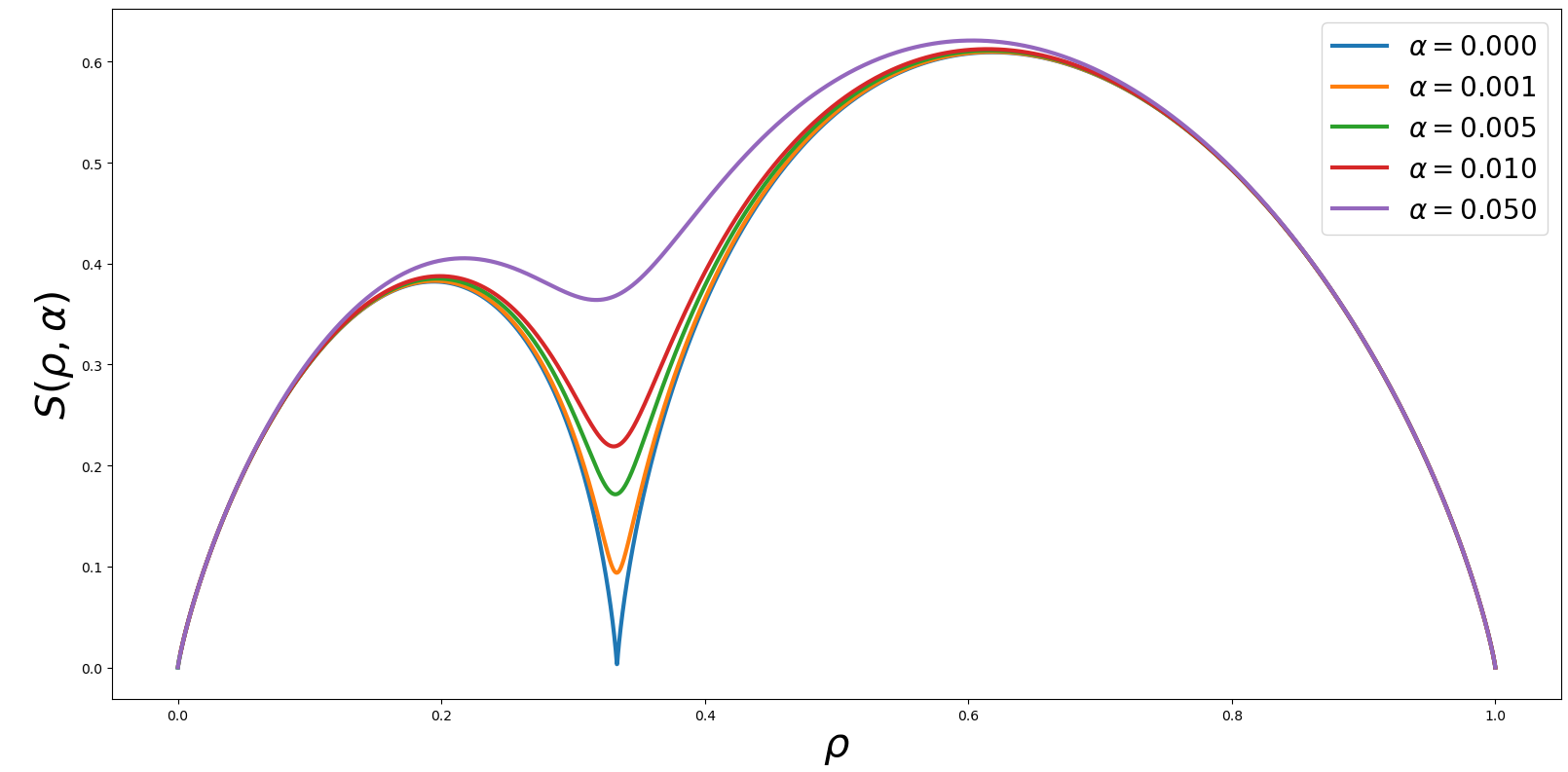}
\end{subfigure}
\caption{The entropy of for the CLG with (a) range $1$, as a function of $\rho$, for various values of $\alpha$. As $\alpha \rightarrow 0$ a cusp develops at the transition point $\rho_c=\frac{1}{2}$, (b) range $2$, as a function of $\rho$, for various values of $\alpha$. As $\alpha \rightarrow 0$ a cusp develops at the transition point $\rho_c=\frac{1}{3}$.}
\label{Splot}
\end{figure*}

\emph{Hyperuniformity} If the initial density $\rho$ is greater than $\frac{1}{n+1}$, and we set $\alpha=0$, the system goes to the active state described in eqn. (\ref{eq:1dclgactss}). However, if $\rho<\frac{1}{n+1}$ and we take the limit of quasistatic driving, $\alpha \rightarrow 0$, the measure concentrates on absorbing states which are on the lowest accessible rung of the transience ladder. These are configurations have, for that $N$ and $L$, the minimal number of $0$s in $0$ clusters of length $>n$, since these have weight $\alpha$ or less. Thus the system tries to put as many $0$s as possible into clusters of length $\le n$, which implies that $0$ clusters must be of length $n$ or greater. The same effect was seen to arise due to repulsion in the lowest density regime in the generalized repulsion processes studied in \cite{krapivsky2013dynamics}. The generating function of the measure on the absorbing states is given by

\beq
C^{abs}(x,y) = \left(1 - y - x y^{n+1} \right)^{-1}.
\eeq
Thus we need to analyse the largest roots of the equation
\beqa
\Lambda^{n+1} &=& \Lambda^n + x, \nonumber\\
\Lambda &=& x^{\frac{1}{n+1}} (1-\Lambda^{-1})^{-\frac{1}{n+1}}. \label{eq:Lambdan}
\eeqa
Now,
\beq
\rho = \frac{x}{\Lambda} \frac{\partial \Lambda}{\partial x} \approx \frac{1}{n+1} - \frac{x^{-\frac{1}{n+1}}}{(n+1)^2} + O(x^{-\frac{2}{n+1}}). \label{eq:xclg}
\eeq
The near-critical limit $\rho \rightarrow \frac{1}{n+1}$ thus corresponds to $x, \Lambda \rightarrow \infty$. The $(n+1)$ roots of eqn. (\ref{eq:Lambdan}), to order $x^{\frac{1}{n+1}}$, lie on a circle in the complex plane with the same radius. The largest root is
\beq
\Lambda_* \approx x^{-\frac{1}{n+1}} \left( 1+\frac{1}{n+1} x^{-\frac{1}{n+1}} +\dots \right). \label{eq:Lstarclg}
\eeq
The roots with the next largest real part are
\beq
\Lambda_\pm =  e^{\frac{\pm 2 \pi i}{n+1}} x^{-\frac{1}{n+1}} \left( 1 + e^{\frac{\mp 2 \pi i}{n+1}} \frac{1}{n+1} x^{-\frac{1}{n+1}} + \dots\right).
\eeq
The correlation length in the system is given in terms of the leading poles of the generating function by
\beq
\xi = \left(\Re \log{ \left(\frac{\Lambda_+}{\Lambda_*} \right)}\right)^{-1} \sim x^{-\frac{1}{n+1}} \sim (\rho-\rho_c)^{-1}.
\eeq

At the critical point, there is only one allowed configuration, $10^n10^n10^n\dots$, which is periodic. Working slightly below the critical point, one expects that on short lengths $l \ll \xi$, the system will look critical, and hence periodic. At length scales $l \gg \xi$, the different parts of the system are uncorrelated with each other. Thus $\xi$ is precisely the hyperuniformity crossover length.

Similar scaling behaviour of the number fluctuations is expected approaching $\rho_c$ from the active side as well. As was shown in \cite{dandekar2013class}, the correlation length in the active state also diverges as $(\rho-\rho_c)^{-1}$. Thus, we have for the conserved lattice gases, for all ranges $n$,
\beq
\nu_h = \nu^{act}_{\perp} = 1.
\eeq

\emph{Entropy cusp at the transition}
Define the configurational entropy per unit length of a system as
\beq
s(\rho) = \lim_{L\rightarrow\infty} \frac{1}{L} \langle -\log{P(C)} \rangle_{\textrm{micro}},
\eeq
where $P(C) = W(C)/Z(N,L)$, where $Z(N,L)$ is the microcanonical partition function. The average is over the ensemble with fixed $L$ and $N = \rho L$. As shown in Appendix B, one can write this entropy in terms of largest pole of the grand partition function,
\beq
s(\rho,\alpha) = \log{\Lambda_*} - \rho \log{(x)} + e \label{eq:entropy},
\eeq
where $e$ the energy density of the system. If $W(C)$ is the weight of the configuration $C$, $e$ is defined as
\beq
e = \lim_{L\rightarrow\infty} \frac{1}{L} \langle - \log{(W(C))} \rangle  = -\log(\alpha) \alpha \frac{\partial \Lambda_*}{\partial \alpha}.
\eeq
The second equality follows from the definition of the weights of the CLG, eqn. (\ref{eq:clgwts}). 

In the limit $\alpha \rightarrow 0$, the system has an equal measure on all allowed states, on both sides of the transition. Thus we can set $e=0$ in eqn. (\ref{eq:entropy}). From eqns. (\ref{eq:xclg}) and (\ref{eq:Lstarclg}), we obtain the entropy of absorbing states slightly below $\rho_c$,
\beq
s^-(\rho,\alpha=0) \approx -(n+1) (\rho_c-\rho) \log{(\rho_c-\rho)}
\eeq
From the generating function for the active steady state, eqn. (\ref{eq:1dclgactss}), it was calculated in \cite{dandekar2013class} that near $\rho_c$,
\beq
(n+1)(\rho-\rho_c) \approx \Lambda_*(\rho) \approx x^{\frac{1}{n+1}}. \nonumber
\eeq
Giving the entropy of near-critical active states to be
\beq
s^+(\rho,\alpha=0) \approx -(n+1) (\rho-\rho_c) \log{(\rho-\rho_c)}.
\eeq
Thus, we see that the entropy goes to $0$ on both sides of the transition. At $\rho = \rho_c$ there is only one allowed configuration, $10^n10^n\dots$, and $s(\rho_c)=0$.

For $\alpha >0$, we can still obtain parametric equations for $\rho(\Lambda)$, $x(\Lambda)$ and $e(\Lambda)$, which are given in Appendix C. There we also calculate that for $\alpha \ll 1$ at $\rho=\frac{1}{n+1}$, the entropy is given by
\beq
s\left(\rho=\rho_c,\alpha\right) = -\frac{1}{n+1}\sqrt{\alpha} \log{\alpha} + O(\log{(\alpha)}).
\eeq
Thus the entropy has a square root cusp along the $\alpha$ direction, for all $n$.

The entropy for the $n=1$ and $n=2$ cases is plotted in fig. \ref{Splot}. We see that $s(\rho)$ is non-monotonic in $\rho$ for a given $\alpha$, with a minimum at $\rho = \rho_c$. The non-monotonicity becomes more pronounced as $\alpha$ is decreased, finally developing into the cusp at $\rho_c$. Thus, not only does the transience field allow us the analyse the measure over the absorbing states in the quasistatic limit, but even for finite values of the field when there are no absorbing states, we can see that the available phase space shrinks for a certain range of $\rho$ and then increases again. Even though the measure for any finite $\alpha$ is an equilibrium measure, it seems the presence of a transience ladder and the incipient active-absorbing phase transition are signalled by this non-monotonicity.\\

\emph{Hyperuniformity in the 1D Abelian Sandpile} The Abelian Sandpile model was one of the earliest models discovered to show self-organised criticality in its driven-dissipative version\cite{bak1987self,dhar1999abelian}. The version with closed boundaries shows an active-absorbing transition as the total number of particles on the lattice is increased. The transition point is known exactly only for some simple classes of graphs, and can depend on the preparation protocol \cite{fey2010driving,fey2010approach}. Due to the deterministic nature of the model, the active steady state exhibits several unusual features, such as ergodicity breaking \cite{dall2006exact}. We study the model below $\rho=1$ on the 1D ring, where there is no active state.

Consider the ASM with $N$ particles on a ring of $L$ sites. The number of particles on the $i^{th}$ site is denoted $n_i$. If $n_i>1$, this site is unstable, and a toppling move occurs (at rate $1$):
\beqa
n_i \rightarrow n_i-2, ~~~ n_{i\pm 1} \rightarrow n_{i\pm 1} + 1.
\eeqa

For the ASM, we introduce a quasistatic drive $\alpha$ such that for all non-empty sites, one particle jumps from the site to a random neighbour at rate $\alpha$. We refer to this as the `$\alpha$ process'. We only work in the quasistatic limit $\alpha \rightarrow 0$, for $\rho <1$, in which case, the $\alpha$ process happens only after all sites have stabilised under the ASM rules. However, if the $\alpha$ hop takes a particle to an already occupied site, this creates an unstable site, leading to an avalanche of ASM topplings, ending with the system reaching another absorbing state. We now determine the steady-state measure on absorbing configuration induced by the $\alpha$ process.

Consider a contiguous cluster of $1$s on the lattice, of length $n$. Say the $i^{th}$ particle in the cluster moves to the $(i+1)^{th}$ site, due to the $\alpha$ process. This makes $(i+1)^{th}$ site unstable, and it topples, transferring one particle to the (now empty) $i^{th}$ site, and one to the $(i+2)^{th}$ site, making this site unstable in turn. This process continues until the avalanche reaches the right end of the cluster. A toppling at the right end of the cluster results in one particle being thrown out of the cluster to the right, and the cluster length reducing by 1. 

Note that the same final state (the cluster length reducing by $1$, one particle thrown out to the right) is reached if any of the $n$ particles in the cluster hops to the right. Similarly, if any of the $n$ particles in the cluster hops to the left, the clusters throws out a particle to the left. Thus the cluster length reduces by $1$ at rate $n \alpha$. 

Thus the quasistatically driven ASM can be mapped to a zero-range process with out-rates $\gamma(k) = k \alpha$ for a site with $k$ particles. (To map an ASM absorbing configuration to a ZRP state, the 0's become sites and the 1-cluster following a 0 becomes the mass on that site.) The steady-state for the ZRP is known exactly, and thus we get the same measure on absorbing configurations for the quasistatically driven ASM:
\beq
W(\{k_1,k_2,\dots\}) = \prod_i \frac{1}{k_i!}, \label{eq:asmwts}
\eeq
for a configuration with $1$-clusters of lengths $k_1, k_2, \dots$. The grand canonical partition function is given by
\beq
C(x,y) = \left( 1- y e^{xy} \right)^{-1}.
\eeq

We again denote $\Lambda = y^{-1}$. The largest pole of $C(x,\Lambda^{-1})$ gives the asymptotic behavior of the canonical partition for $L$ sites, for large $L$. As shown in Appendix D, the poles of the generating function are given by
\beq
\Lambda_k = \frac{x}{W_k(x)}, \mbox{ with } x = \frac{\rho}{1-\rho} e^{\frac{\rho}{1-\rho}}. \label{eq:lambdaasm}
\eeq
where $W_k$ is the $k^{th}$ branch of the Lambert function $W(x)$, and $k$ is an integer in the range $-\infty$ to $\infty$. The Lambert function \cite{veberic2010having} is defined though,
\beq
W(x) e^{W(x)} = x,
\eeq
which has an infinity of roots in the complex plane, labeled by $k$.

Now, the ASM at $\rho=1$ has a single absorbing state, $1111\dots$, and is trivially periodic. As the transition point at $\rho=1$ is approached, we expect the correlation length of the system to diverge. The leading pole is given the real branch of eqn. (\ref{eq:lambdaasm}), $\Lambda_0(x) = x/W_0(x)$. The next largest roots of the denominator of the generating function are the pair of conjugate roots with $k = \pm 1$. Now, we have for large $y$,
\beq
\frac{\Lambda_{\pm 1}(ye^y)}{\Lambda_0(ye^y)} = \frac{W_0(ye^y)}{W_{\pm 1}(y e^y)} \approx 1 \pm \frac{2 \pi i}{y} - \frac{2 \pi (\pm i + 2\pi)}{y^2} + O(y^{-3}),
\eeq
Since $x = \frac{\rho}{1-\rho} e^{\frac{\rho}{1-\rho}}$, we have that the correlation length near $\rho = 1$ is
\beq
\xi = \left(\Re\log{\left(\frac{\Lambda_{\pm 1}} {\Lambda_0}\right)}\right)^{-1} \approx \frac{1}{4\pi^2(1-\rho)^2},
\eeq
while over shorter distances the system is periodic with a periodicity given the imaginary part $\sim (1-\rho)^{-1}$. Hence, the hyperuniformity correlation exponent is $\nu_h = 2$. As the system is periodic on length scales $\ll \xi$, we have that the hyperuniformity exponent $\alpha=1$.




\emph{Entropy of Absorbing States} Since the distribution of gaps between $0$s, eqn. \ref{eq:asmwts}, is a Bernoulli measure, we know that in the grand canonical emsemble this splits into independent Poisson measures for each individual gap. Near $\rho=1$, each gap is Poisson-distributed with mean $(1-\rho)^{-1}$. Thus the total entropy of the system is 
\beq
L s(\rho) = N_0 s_{\textrm{Poisson}}((1-\rho)^{-1}),
\eeq
where $N_0$ is the number of $0$s in the system, and $s_{\textrm{Poisson}}(\lambda)$ is the entropy of a Poisson distribution with mean $\lambda$. Although no closed form expression exists for $s_{\textrm{Poisson}}$, in the limit of large $\lambda$, it can be written as a series in $\lambda^{-1}$,
\beq
s_{\textrm{Poisson}}(\lambda) = \frac{1}{2} \log{(2\pi e \lambda)} - \frac{1}{12\lambda} + O(\lambda^{-2}),
\eeq
Hence, we have that,
\beq
s(\rho) = -\frac{1}{2} (1-\rho)\log{(2\pi (1-\rho)e)} - \frac{1}{12} (1-\rho)^2 + O((1-\rho)^3).
\eeq

Thus the entropy goes to $0$ at $\rho_c$ as $\frac{1}{2} \delta \rho \log{(\delta \rho)}$. A different method for calculating the entropy is given in Appendix E.\\

\emph{Conclusions} We have shown that for the class of longer-ranged Conserved Lattice Gases defined in \cite{dandekar2013class}, and the fixed energy Abelian Sandpile Model in 1D, an exactly calculable measure over the absorbing states can be defined using an external field to induce quasistatic driving. We calculated the hyperuniformity exponents for both cases and showed that the entropy goes to zero as $\rho \rightarrow \rho_c$. For the Conserved Lattice Gases, we could calculate the measure exactly even for a finite external field, and showed that the entropy of configurations shows nonmonotonic behaviour around the zero-field transition point, developing into a cusp as the drive is tuned to zero.

The calculation of hyperuniformity in the ASM offers a concrete example of a model where a nontrivial hyperuniformity exponent ($\nu_h=2$) can be calculated exactly, and is the first known result of this sort, to our knowledge. 

For the CLGs, the non-monotonic behaviour of the entropy even for finite $\alpha$ where there are no transient states offers an understanding of the entropy cusp seen in studies of active-absorbing transitions \cite{martiniani2019quantifying} is not dependent on the exact nature of the measure on the absorbing states (and hence the particular initial conditions employed) but originates simply in the squeezing of the available phase space, seen even in the driven system which has no absorbing states. As the drive is taken to zero, for $\rho$ values before the cusp, this driven measure concentrates on absorbing states, while after the cusp it concentrates on active states. Thus, we believe this allows us to understand active-absorbing phase transitions generically in terms simply of the relative \emph{available phase space} for absorbing and active states, which switches from being in favour of one to the other at $\rho_c$. Since active-absorbing transitions break no symmetries, an thermodynamic understanding of their origin is lacking, and the transience field provides a potential route to the resolution of such an origin.\\

\emph{Acknowledgements} I would like to thank Deepak Dhar for suggesting the idea of the slow drive in the ASM, and his constant encouragement and guidance.

\bibliographystyle{unsrt}
\bibliography{hyperu}

\clearpage
\newpage

\begin{widetext}
\appendix

\section{The grand canonical partition function}

The microcanonical partition function $Z(N,L)$ is defined as the sum of the weights of all configurations $C$ on a ring of $L$ sites, with $N$ particles, with $n_i(C)$ denoting the number of particles on site $i$ in configuration $C$,

\beq
Z(N,L) = \sum_C W(C) \delta\left(\sum_i n_i(C) - N\right).
\eeq

For a lattice of $0$ sites we set $Z(N,0)= \delta_{N,0}$ by definition. The canonical partition function, is defined for a lattice of fixed size $L$, while the number of particles is allowed to vary:
\beq
Z_L(x) = \sum_{N=0}^{\infty} x^N Z(N,L).
\eeq

The grand canonical partition is defined as
\beq
C(x,y) = \sum_{L=0}^{\infty} \sum_{N=0}^{\infty} x^N y^L Z(N,L).
\eeq
Let us suppose that the weights $W(C)$ break up into products of weights of subconfigurations, as happens for the models studied in this paper. Consider the minimal set of subconfigurations needed to define the weights in this fashions. This implies that every given configuration can be formed from the minimal set in a unique way. For example, in the conserved lattice gas with $\alpha=0$ and $\rho>\frac{1}{2}$, the minimal set is simply $1$ and $10$. Denote the minimal set of subconfigurations by $\{ c_1, c_2, \dots \}$, and let $n_i$ and $l_i$ denote the number of particles and length respectively of subconfiguration $c_i$. We define
\beq
c(x,y) = \sum_{i} c_i x^{n_i} y^{l_i}.
\eeq
By using the fact that the configurations summed over in $C(x,y)$ are sequences of all allowed subconfigurations with the correct weights, one has
\beq
C(x,y) = \frac{a(x,y)}{1-c(x,y)}.
\eeq
where $a(x,y)$ depends only on the boundary condition. Consider the poles of $C(x,y)$, which are the roots of $c(x,y)$,
\beq
c(x,y)=1 \mbox{ at } y= y_1(x), y_2(x), y_3(x), \dots
\eeq
Then,
\beqa
C(x,y) = \sum_{L=0}^{\infty} y^L \left(a_1(x) (y_1(x))^{-L} + a_2(x) (y_1(x))^{-L} + \dots \right).
\eeqa
For convenience of notation we denote $y^{-1}$ as $\Lambda$, and this implies
\beq
Z_L(x) = \Lambda_1(x)^L + \Lambda_2(x)^L + \Lambda_3(x)^L + \dots.
\eeq
Let $\Lambda_1 \ge \Lambda_2 \ge \Lambda_3 \ge \dots$. Then, we have
\beq
\lim_{L\rightarrow\infty} \frac{1}{L}\log Z_L(x) = \log{\Lambda_1(x)}.
\eeq
And the subleading correction to this `free energy' is 
\beq
C \left(\frac{\Lambda_2}{\Lambda_1}\right)^L \sim C \exp{\left(-\frac{L}{\xi}\right)},
\eeq
where $\xi$ is the largest correlation length in the system, and which is thus given by
\beq
\xi = \left(\log{\left(\frac{\Lambda_2}{\Lambda_1}\right)} \right)^{-1}.
\eeq


\section{Configurational Entropy from the partition function}

The canonical partition function $Z_L(x)$ can be written as
\beq
Z_L(x) = \sum_{N=0}^{\infty} \exp{\left(\log{Z(N,L)} + N \log{(x)}\right)}. \label{eq:b1}
\eeq
Changing variables to $\rho = N/L$ and $L$, converting the sum to an integral, and using a saddle point approximation, and the typical value of $\rho$ in the system is
\beq
\log{x} = -\frac{\partial}{\partial \rho} Z(\rho^* L,L).
\eeq
If the saddle-point approximation is valid, $\rho^*$ is also the mean density of the system with fugacity $x$,
\beq
\rho^* = \frac{x}{L} \frac{\partial}{\partial x}\log{Z_L(x)} = x\frac{\partial}{\partial x}\log{\Lambda_1(x)},
\eeq
Taking logarithms of both sides of eqn. (\ref{eq:b1}),
\beq
\lim_{L\rightarrow\infty} \frac{1}{L} \log{Z_L(x)} = \lim_{L\rightarrow\infty} \frac{1}{L} \log{Z(\rho^* L,L)} + \rho^* \log(x).
\eeq
Now, for a system with fixed $N$ and $L$, the probabilities of various configurations are given as $P(C) = W(C)/Z(N,L)$. The configurational entropy is
\beq
S(N,L) = -\sum_{C} P(C) \log{P(C)} = \langle -\log{W(C)} \rangle + \log{Z(N,L)}.
\eeq
Defining the energy density $e = \frac{1}{L} \langle -\log{W(C)} \rangle$, and the entropy density $s = \frac{S}{L}$, we have
\beq
\lim_{L\rightarrow \infty} \frac{1}{L} \log{Z(N,L)} = s(\rho) - e(\rho) .
\eeq
Therefore, we have
\beq
s(\rho^*) = \log{\Lambda_1(x)} - \rho^* \log{x} + e(\rho^*).
\eeq

\section{Parametric expressions for $\rho$, $x$ and $e$ for the CLG for general $n$ and $\alpha$}

\beqa
\rho(\Lambda,\alpha,n) &=& \frac{(\Lambda -1) (\Lambda -\alpha ) \left((\alpha -1) \Lambda +(\alpha +\Lambda )    \Lambda ^{n+1}\right)}{\Lambda  \left((\alpha -1) \left(-\alpha +\Lambda^2+(\Lambda -1) (n+1) (\Lambda -\alpha )\right)+\left(2 \alpha  \Lambda -\alpha (\alpha +2)+\Lambda ^2\right) \Lambda ^{n+1}\right)},\\
x(\Lambda,\alpha,n) &=& \frac{-(\alpha +1) \Lambda ^{n+1}+\alpha \Lambda ^n+\Lambda ^{n+2}}{\alpha +\alpha \Lambda ^n+\Lambda ^{n+1}-1},\\
e(\Lambda,\alpha,n) &=& \frac{\alpha  (\Lambda -1) \log (\alpha ) \left(\Lambda +2 \Lambda^{n+1}-1\right)}{(1-\alpha ) \left(\alpha -\Lambda ^2+(\Lambda -1) (n+1) (\alpha -\Lambda )\right)+\left(2 \alpha  \Lambda -\alpha  (\alpha +2)+\Lambda ^2\right) \Lambda ^{n+1}}.
\eeqa

To get the behaviour of $\Lambda$ at the critical point, we set $\rho=\frac{1}{n+1}$ in the first equation and solve for $\alpha$. Expanding this for $\Lambda \ll 1$ (which is the correct limit near the critical point), as for $\alpha=0$, $\Lambda=0$ at $\rho=\rho_c$ in the active steady state \cite{dandekar2013class}), we get

\beqa
\alpha\vert_{\rho=\rho_c} \approx \Lambda^2 \mbox{ for all $n$},\\
x(\Lambda,\Lambda^2,n) \approx \Lambda^{n+1},\\
e(\Lambda,\Lambda^2,n) \approx -\frac{2}{n+1}\Lambda \log{\Lambda}.
\eeqa 
Putting this into eqn. \ref{eq:entropy} gives
\beq
s(\rho=\rho_c) = -\frac{2}{n+1}\Lambda \log{\Lambda} + O(\Lambda) \approx -\frac{1}{n+1}\sqrt{\alpha} \log{\alpha}
\eeq

\section{Poles of the generating function for the 1D ASM}

The grand canonical partition function for the 1D ASM is given by
\beq
C(x,y) = \left( 1- y e^{xy} \right)^{-1}.
\eeq

We again denote $\Lambda = y^{-1}$. The largest pole of $C(x,\Lambda^{-1})$ gives the asymptotic behavior of the canonical partition for $L$ sites, for large $L$. The equation for the poles can be solved in terms of the Lambert function $W(x)$,
\beqa
1- e^{x/\Lambda}/\Lambda &=& 0.\\
\mbox{Thus, }\frac{x}{\Lambda}e^{x/\Lambda} &=& x,\\
\mbox{Giving }\Lambda &=& \frac{x}{W(x)}.
\eeqa

Where $W(x)$ is the Lambert function, defined implicitly through
\beq
W(x) e^{W(x)} = x.
\eeq
The equation thus has an infinity of roots, given by all the branches of the Lambert function in the complex plane. The density is related to the fugacity and the partition function via
\beqa
\rho &=& x \frac{d}{dx} \log{(\Lambda(x))}\\
&=& 1 - \frac{1}{1+W(x)}.
\eeqa
The last equality follows from the definition of the Lambert function, and is valid for the real branch $W_0(x)$ and hence the root $\Lambda_0(x)$, which is also the root with the lartgest real part. Thus, we get 
\beq
W(x) = \frac{\rho}{1-\rho}.
\eeq
Which gives
\beqa
x &=& \frac{\rho}{1-\rho} e^{\frac{\rho}{1-\rho}},\\
\mbox{and }\Lambda_0(\rho) &=& e^{\frac{\rho}{1-\rho}}.
\eeqa

\section{Entropy of absorbing states for the 1D ASM}

Knowing the dependence of the largest pole of grand canonical partition function, and the fugacity $x$ on density, one can determine the entropy of absorbing states from eqn. (\ref{eq:entropy}). However, since we determine the measure only in the limit $\alpha \rightarrow 0$, the function $e(\rho)$ has to be determined using a different method. We define a parameter $\beta$ by generalising the weights in eqn. (\ref{eq:asmwts}) to
\beq
W_{\beta}(\{k_1,k_2,\dots\}) = \prod_i \left(\frac{1}{k_i!}\right)^{\beta} \label{eq:asmwtsbeta}
\eeq
for a configuration with $1$-clusters of lengths $k_1, k_2, \dots$. The ASM measure is the limit $\beta=1$. The grand canonical partition function for general $\beta$ can we written as
\beq
C_{\beta}(x,y) = \left( 1- y e_{\beta}^{xy} \right),
\eeq
where we have defined
\beq
e_{\beta}^{z} \equiv \sum_{n=0}^{\infty} \left(\frac{1}{n!}\right)^{\beta} z^n.
\eeq
Denote by $W_\beta$ the leading solution to the equation $W_\beta = e^{W_\beta}_\beta$. From the definition of the energy, we have
\beq
e(\rho) = -\frac{\partial F}{\partial \beta}\vert_{\beta=1}
\eeq
Although the properties of the generalized exponential $e_\beta(x)$ are not expressible in a closed form, we only need the behaviour near $\beta=1$ and near the transition point $\rho=1$. We use a saddle-point expansion to approximate the sum over $k$, and Stirling's approximation for the factorials $k!$ appearing in the sum. The final result for the energy density is that
\beq
e(\rho) = -\rho\left(1- \log{\left(\frac{\rho}{1-\rho}\right)}\right) + \frac{1}{2} (1-\rho) \log{\left(2 \pi e \frac{\rho}{1-\rho}\right)} + O((1-\rho)^2).
\eeq
And hence
\beq
s(\rho) = \frac{1}{2} (1-\rho)\log{(1-\rho)} + O((1-\rho)).
\eeq
\end{widetext}

\clearpage

\end{document}